\documentstyle[11pt,epsfig]{article}
\title{\bf A fundamental length as a candidate for dark energy:
\\a DSR inspired FRW spacetime}
\author{N. Khosravi\thanks{email: n-khosravi@sbu.ac.ir, Telephone: +98 (21) 29902796, Fax: +98 (21) 22431666.}\
and H. R. Sepangi\thanks{email: hr-sepangi@sbu.ac.ir}\\ {\small
Department of Physics, Shahid Beheshti University, Evin, Tehran
19839, Iran}}

\begin{document}
\maketitle
\begin{abstract}
We show that the existence of a fundamental length, introduced in
Deformed Special Relativity (DSR) inspired minisuper (phase-) space,
causes the behavior of the scale factor of the universe to change
from that of a universe filled with dust to an accelerating universe
driven by a cosmological constant.

\noindent\\
PACS: 98.80.-k; 95.36.+x; 06.20.Jr.
\\Keywords: Cosmology; Dark energy; Fundamental
length.
\end{abstract}

\maketitle
\section{Introduction} The question of the evolution of
cosmos from the beginning to its present stage constitutes one of
the most important questions in the history of science. More
precisely, the question may be asked as to how the present
accelerating phase of the universe could result from the big-bang.
To answer this question, a heap of observational data have been
collected and a huge amount of theoretical work has been done. To
describe the behavior of the present accelerating phase of the
universe discovered a few years ago, numerous models have been
devised and introduced \cite{acceleration}. A common tool used to
describe the accelerated expansion of our universe is what is
known as dark energy which is the basic ingredient in model
theories with a cosmological constant.

It is generally believed that quantization of gravity could lead to
the removal of singularities which one encounters in classical
general relativity. On the other hand it is also generally believed
that the existence of a fundamental length is a natural feature in
all theories that endeavor to answer the old and interesting
question of how to quantize gravity \cite{gar,rovelli}. This
fundamental length has been introduced in some model theories by
hand \cite{dsr111,dsr1,dsr11,dsr2,dsr3,dsr4,ncgeometry,ncg1,ncg2}
where it is shown that such effective theories can be recovered as
the limit of full quantum gravity \cite{dsr111,dsr1,amelino,a1,a2}.
Therefore, the motivation behind the construction of these effective
models is to study the effects of such a fundamental length scale on
simple scenarios which are exactly solvable. One of these approaches
is to modify or deform the algebraic structure of the phase-space.
Such deformations may be done in various manners, a few example of
which can be found in \cite{ncfields,nc1,nc2,nc3,babak,b1,nima}.

To study the effects of the existence of a fundamental length in a
cosmological scenario, one can construct a model based on the
noncommutative structure of the Deformed (Doubly) Special Relativity
\cite{dsr1} which is related to what is known as the
$\kappa$-deformation \cite{gilkman,gli1}. This way of introducing
noncommutativity is interesting because of its compatibility with
Lorentz symmetry \cite{gilkman,gli1,romero}. The
$\kappa$-deformation is introduced and studied in
\cite{ruegg,r1,shahn}. The $\kappa$-Minkowski space
\cite{shahn,freidel} arises naturally from the $\kappa$-Poincare
algebra \cite{ruegg,r1} such that the ordinary brackets between the
coordinates are replaced by
\begin{eqnarray}\label{kappa-nc}
\{{x}_0,{x}_i\}=\frac{1}{\kappa}{x}_i,
\end{eqnarray}
where $\{,\}$ represents the Poison bracket and $\kappa$ is the
deformation (noncommutative) parameter which has the dimension of
mass $\kappa=\epsilon \ell^{-1}$ when $c=\hbar=1$, where
$\epsilon=\pm1$ \cite{bruno} such that $\kappa$ and $\ell$ are
interpreted as dimensional parameters that identify with the
fundamental energy and length, respectively. In the following we
restrict ourselves to the sector $\epsilon=-1$. This fundamental
length can be identified, for example, with the Planck length. As
mentioned above, one can change the structure of the phase-space
based on equation (\ref{kappa-nc}). To study the effects of this
kind of deformation, we briefly review the Hamiltonian formalism of
the FRW spacetimes in the next section. In section \ref{3}, the
$\kappa$-deformed phase-space for FRW spacetimes is introduced and
its effects are studied. The paper ends with a discussion of the
results.

\section{Phase-space structure of the FRW spacetimes}
Let us start by briefly studying the ordinary FRW model
\begin{eqnarray}\label{de-sitter}
ds^2=-N^2(t)dt^2+a^2(t)(dx^2+dy^2+dz^2),
\end{eqnarray}
where $N(t)$ is the lapse function. The Einstein-Hilbert
Lagrangian with  a general energy density $V(a)$ becomes
\begin{eqnarray}\label{lagrangian}
{\cal{L}}&=&\sqrt{-g}(R[g]-V(a))\nonumber\\
&=&-6N^{-1}a\dot{a}^2- Na^3 V(a),
\end{eqnarray}
where $R[g]$ is the Ricci scalar and in the second line the total
derivative term has been ignored. The corresponding Hamiltonian up
to a sign becomes
\begin{eqnarray}\label{hamiltonian1}
{\cal{H}}_0&=&\frac{1}{24}Na^{-1}p_a^2- Na^3V(a).
\end{eqnarray}
Here, we note that since the momentum conjugate to $N(t)$,
$\pi=\frac{\partial{\cal{L}}}{\partial \dot{N}}$ vanishes, the
term $\lambda \pi$ must be added as a constraint to Hamiltonian
(\ref{hamiltonian1}). The Dirac Hamiltonian then becomes
\begin{eqnarray}\label{hamiltonian}
{\cal{H}}&=&\frac{1}{24}Na^{-1}p_a^2- Na^3V(a)+\lambda \pi.
\end{eqnarray}
Now, the equations of motion with respect to the above Hamiltonian
are given by
\begin{eqnarray}\label{eqofmotion}
\dot{a}&=&\left\{a,{\cal{H}}\right\}=\frac{1}{12}Na^{-1}p_a,\nonumber\\
\dot{p_a}&=&\left\{p_a,{\cal{H}}\right\}=\frac{1}{24}Na^{-2}p_a^2+3
N a^2V(a)+N a^3 V'(a),\nonumber\\
\dot{N}&=&\left\{N,{\cal{H}}\right\}=\lambda,\nonumber\\
\dot{\pi}&=&\left\{\pi,{\cal{H}}\right\}=-\frac{1}{24}a^{-1}p_a^2+a^3
V(a),\label{constraint}
\end{eqnarray}
where a prime represents differentiation with respect to the
argument. Note that to satisfy the constraint $\pi=0$ at all times
the secondary constraint $\dot{\pi}=0$ must also be satisfied. A
simple calculation leads to
\begin{eqnarray}\label{eqofmotion1}
\dot{a}=\sqrt{\frac{1}{6} a^2 V(a)},\label{1equation}
\end{eqnarray}
where in the above we fix the gauge by taking $N=1$, that is, we
work in the comoving gauge. Note that the other equations will
automatically become consistent if the above equation is
satisfied. The solution for dust for which $V(a)=\rho_0 a^{-3}$
becomes
\begin{eqnarray}\label{com.solutions}
a(t)=\frac{3^{\frac{1}{3}}}{2}\left(\sqrt{\rho_0}\hspace{2mm}t+C_1\right)^{\frac{2}{3}},
\end{eqnarray}
where $C_1$ is the integration constant which is set to zero. This
scale factor shows a singularity at $t=0$, the so called big-bang
singularity, and has a power-law behavior which is not consistent
with late time observations.

\section{$\kappa$-deformed phase-space structure of the FRW
spacetimes}\label{3} It has long been argued that the deformation in
phase-space can be seen as an alternative path to quantization,
based on Wigner quasi-distribution function and Weyl correspondence
between quantum-mechanical operators in Hilbert space and ordinary
c-number functions in phase space, see for example
\cite{quantization} and the references therein. The deformation in
the usual phase-space structure is introduced by Moyal brackets
which are based on the Moyal product \cite{ncfields,nc1,nc2,nc3}.
However, to introduce such deformations it is more convenient to
work with Poisson brackets rather than Moyal brackets.

From a cosmological point of view, models are built in a
minisuper-(phase)-space. It is therefore safe to say that studying
such a space in the presence of the deformations mentioned above can
be interpreted as studying the quantum effects on cosmological
solutions. One should note that in gravity (here, cosmology) the
effects of quantization are woven into the existence of a
fundamental length \cite{gar}, as mentioned in the introduction. The
question then arises as to what form of deformations in phase-space
is appropriate for studying quantum effects in a cosmological model?
Studies in noncommutative geometry \cite{ncgeometry,ncg1,ncg2} and
generalized uncertainty principle (GUP) \cite{kempf} have been a
source of inspiration to answer the above question. More precisely,
introduction of modifications in the structure of geometry in the
way of noncommutativity \cite{ncgeometry,ncg1,ncg2} has become the
basis from which similar modifications in phase-space have been
inspired. In this approach, the fields and their conjugate momenta
play the role of coordinate basis in noncommutative geometry
\cite{carmona,c1}. In doing so an effective model is constructed
whose validity will depend on its power of prediction. For example,
if in a model field theory the fields are taken as noncommutative,
as has been done in \cite{carmona,c1}, the resulting effective
theory predicts the same Lorentz violation as a field theory in
which the coordinates are considered as noncommutative
\cite{carlson,car1,car2}. Over the years, a large number of works on
noncommutative fields \cite{ncfields,nc1,nc2,nc3} have been inspired
by noncommutative geometry model theories
\cite{ncgeometry,ncg1,ncg2}. As a further example, it is well known
that string theory can be used to suggest a modification in the
bracket structure of coordinates, also known as GUP \cite{kempf}
which is used to modify the phase-space structure \cite{babak}.

In this paper we will examine a new kind of modification in the
phase-space structure inspired by relation (\ref{kappa-nc}), much
the same as what has been done in \cite{ncfields,babak,nima}.  In
what follows we introduce noncommutativity based on
$\kappa$-Minkowskian space and study its consequences on the
solutions discussed in the previous section. To introduce
noncommutativity one can start with
\begin{eqnarray}\label{kappa-nc-filds}
\{\hat{N}(t),\hat{a}(t)\}= -\ell \hat{a}(t).
\end{eqnarray}
This is similar to equation (\ref{kappa-nc}) since one can interpret
$N(t)$ and $a(t)$, appearing as the coefficients of $dt$ and
$d\vec{x}$, in the same manner as ${x}_0$ and ${x}_i$ appearing in
(\ref{kappa-nc}) respectively. For this reason we shall call it the
$\kappa$-Minkowskian-minisuper-phase-space. In this case the
Hamiltonian  becomes
\begin{eqnarray}\label{primedhamiltonian}
\hat{{\cal{H}}}_0&=&\frac{1}{24}\hat{N}\hat{a}^{-1}{\hat{p}}_a^2-
\hat{N}{\hat{a}}^3V(\hat{a}),
\end{eqnarray}
where the ordinary Poisson brackets are satisfied except in
(\ref{kappa-nc-filds}). To progress further, one introduces the
following variables \cite{chai}
\begin{eqnarray}\label{newvariables}
\left\{
\begin{array}{ll}
\hat{N}(t)=N(t)+\ell a(t) p_a(t),\\
\hat{a}(t)=a(t),\\
\hat{p}_a(t)=p_a(t).
\end{array}\right.
\end{eqnarray}
It can be easily checked that the above variables satisfy
(\ref{kappa-nc-filds}) if the unprimed variables satisfy ordinary
Poisson brackets. With the above transformations, Hamiltonian
(\ref{primedhamiltonian}) changes to
\begin{eqnarray}\label{hamiltonianNC1}
{\cal{H}}^{nc}_0&=&\frac{1}{24}Na^{-1}{p}_{a}^2-N{a}^3V(a)+\frac{1}{24}\ell
p_a^3-\ell  a^4 V(a) p_a.
\end{eqnarray}
Obviously, the momentum conjugate to $N(t)$ does not appear in the
above Hamiltonian, that is,  $\pi$ is a primary constraint. It can
be checked by using Legendre transformations that the conjugate
momentum corresponding to $N(t)$,
$\pi=\frac{\partial{\cal{L}}}{\partial \dot{N}}$, vanishes. It is
therefore necessary to add the term $\lambda\pi$ to Hamiltonian
(\ref{hamiltonianNC1}) to obtain the Dirac Hamiltonian
\begin{eqnarray}\label{hamiltonianNC}
{\cal{H}}^{nc}&=&\frac{1}{24}Na^{-1}{p}_{a}^2-N{a}^3V(a)+\frac{1}{24}\ell
p_a^3-\ell  a^4 V(a) p_a+\lambda\pi.
\end{eqnarray}
The equations of motion with respect to Hamiltonian
(\ref{hamiltonianNC}) are
\begin{eqnarray}\label{NCeqofmotion}
\dot{a}&=&\left\{a,{\cal{H}}^{nc}\right\}=\frac{1}{12}Na^{-1}p_a+
\frac{1}{8}\ell p_a^2-\ell a^4V(a),\nonumber\\
\dot{p_a}&=&\left\{p_a,{\cal{H}}^{nc}\right\}=\frac{1}{24}Na^{-2}p_a^2+3N
a^2V(a)+Na^3V'(a)+4\ell a^3V(a)p_a+\ell a^4V'(a)p_a,\nonumber\\
\dot{N}&=&\left\{N,{\cal{H}}^{nc}\right\}=\lambda,\nonumber\\
\dot{\pi}&=&\left\{\pi,{\cal{H}}^{nc}\right\}=-\frac{1}{24}a^{-1}p_a^2+a^3V(a),\label{NCconstraint}
\end{eqnarray}
where a prime denotes differentiation with respect to the
argument. Again, we restrict ourselves to the comoving gauge for
which $N=1$. We also note that the secondary constraint,
$\dot{\pi}=0$, must be satisfied in order that the primary
constraint, namely $\pi=0$, is satisfied at all times. If $p_a$ is
now calculated from the secondary constraint, $\dot{\pi}=0$, and
the result is substituted in the first equation in
(\ref{NCeqofmotion}) one obtains
\begin{eqnarray}\label{equationNC}
\dot{a}-2\ell a^4V(a)=\sqrt{\frac{1}{6}a^2V(a)}.
\end{eqnarray}
It is easy to check that the above equation is consistent with the
second equation in (\ref{NCeqofmotion}) as well. Note that this
equation reduces to the commutative case (\ref{1equation}) when
$\ell\rightarrow 0$. The exact solution for $V(a)=\rho_0 a^{-3}$,
representing dust, becomes
\begin{eqnarray}\label{nc.solutions}
a(t)=\frac{1}{2\times3^{\frac{1}{3}}}\left[\frac{-1+  {C_2}
e^{3\ell\rho_0\vspace{2mm}t}}{\ell\sqrt{\rho_0}}\right]^{\frac{2}{3}},
\end{eqnarray}
where $C_2$ is an integration constant which must satisfy $ {C_2}
e^{3\ell\rho_0\vspace{2mm}t}\geq1$. In the following we set $C_2=1$.
We note that the limit $\ell\rightarrow0$ still coincides with the
commutative result (\ref{com.solutions}), predicting a big-bang
singularity. The scale factor (\ref{nc.solutions}) with the chosen
integration constant in the limit $\ell\rightarrow0$ then leads to
\begin{eqnarray}\label{nclimit.solutions}
a(t)=\frac{3^{\frac{1}{3}}}{2}\left(\sqrt{\rho_0}\hspace{2mm}t+{\cal{O}(\ell)}\right)^{\frac{2}{3}}.
\end{eqnarray}
This means that the scale factor begins from a big-bang
singularity and then behaves as if the universe is filled with
dust. However, the difference between the commutative and
noncommutative  cases are manifest in the late time behavior where
in the noncommutative case with $t\rightarrow \infty$, the scale
factor becomes
\begin{eqnarray}\label{cosmologicalconstant}
a(t)\propto e^{2\ell\rho_0\vspace{2mm}t},
\end{eqnarray}
for which (\ref{de-sitter}) represents a de-Sitter metric with a
cosmological constant $\Lambda=12\ell^2\rho_0^2$. Therefore, for
late times the behavior of the scale factor becomes exponential,
pointing to the existence of a fundamental length which can be
interpreted as a candidate for dark energy. Taking the relation
for $\Lambda$ given above, the value of the cosmological constant
can be estimated. If one chooses $\ell=\ell_P$ where
$\ell_P\sim10^{-35}m$ is the Planck length and the matter density
is taken as $\rho_0=1.9\times10^{-29}h^2g cm^{-3}$
\cite{cosmology} where $h$ is the Planck constant, the
cosmological constant is found to be $\Lambda\sim10^{-170}m^{-2}$
which is well within the suggested upper bound
$\Lambda_{upperbound}\sim10^{-52}m^{-2}$, a generally accepted
value determined by observation \cite{wikipedia}. Note that the
value of the cosmological constant in our model is much smaller
than the suggested upper bound. This value could be improved to
become more in line with the observational upper bound if other
factors such as radiation are taken into account. This means that
the scale factor begins with a dust-like behavior before entering
the accelerating phase (cosmological constant domination) which is
a direct consequence of the existence of a fundamental length.

\section{Discussion}\label{4}
The construction of an effective theory is usually the result of
the observation that the the corresponding full theory is too
complicated to deal with. This, for example, is true in describing
the quantum effects on cosmology since the full theory is
immensely difficult to handle \cite{martin}. The DSR can be
interpreted as one such effective theory. In the present study, we
have introduced a fundamental length by employing an effective
theory, namely a DSR-inspired model. In doing so we have chosen
the simplest cosmological model, namely a FRW spacetime in the
presence of matter in the form of dust. Here, the introduction of
a fundamental length causes additional terms to appear in
Hamiltonian (\ref{hamiltonianNC}) as compared to
(\ref{hamiltonian}). As has been mentioned in
\cite{nima,quantization,chern}, these extra terms can be
interpreted as the effects of high energy corrections of a full
theory, e.g., the string theory\footnote{This interpretation is
consistent with the one introduced at the beginning of section
\ref{3} in that the extra terms can be interpreted as quantum
effects.}. On the other hand, our results can be used to address
one of the problems of interest in cosmology, i.e. dark energy. It
is therefore reasonable to assume that the modification introduced
through relation (\ref{kappa-nc-filds}) may also be relevant in
model theories dealing with scenarios involving dark energy and
the accelerating universe.

In this paper, we have shown that the existence of a fundamental
length can play the role of a cosmological constant at late times.
This means that such a notion can be employed to account for dark
energy at late times. Note that the cosmological constant is only a
parameter which is introduced by hand to describe the accelerating
behavior of the universe. Here however, its appearance at late times
is the direct consequence of the existence of the fundamental
length. Although we have introduced such a length by hand here, its
existence has its roots in quantum gravity \cite{gar}, considered as
the full theory. It therefore seems plausible that the existence of
a fundamental length can describe the present accelerating behavior
of the universe.

\vspace{5mm}\noindent\\
\textbf{Acknowledgments}\vspace{2mm}\noindent\\
We would like to thank  M. M. Sheikh-Jabbari  and M. Bojowald for
useful discussions. We also thank the anonymous referee for
useful comments.

\end{document}